\documentclass[11pt]{article}
\usepackage{graphicx}
\date{}
\usepackage[latin1]{inputenc}
\usepackage{amsmath}
\usepackage{amsfonts}
\usepackage{amssymb}
\begin{document}
\title{\textbf{Nonlinear sigma model of a spin ladder containing a static single hole}}
\author{{A. R. Pereira}$^{1,\,2}$ \thanks{E-Mail:
apereira@ufv.br}, {E.
Ercolessi}$^{2,3}$\thanks{E-Mail:Elisa.Ercolessi@bo.infn.it}, A.S.T.
Pires$^{4,\,}$ \thanks{E-mail: antpires@fisica.ufmg.br}\\ 
\small $^{1}$ \it
Departamento de F\'{\i}sica, Universidade Federal de Vi\c{c}osa\\
\small \it 36570-000, Vi\c{c}osa, Minas
Gerais, Brazil\\
\small$^{2}$\it Physics Department, University of Bologna, Via
Irnerio 46, I-40126, Bologna, Italy\\
\small $^3$INFN and CNISM, Bologna, Italy\\
\small $^{4}$ \it Departamento de F\'{\i}sica, ICEX, Universidade
Federal de Minas Gerais\\ \small \it  Caixa Postal 702, 30123-970,
Belo Horizonte, Minas Gerais, Brazil}

\maketitle
\begin{center}
\textbf{Abstract}
\end{center}

\indent In this letter we extend the nonlinear $\sigma$ model
describing pure spin ladders with an arbitrary number of legs to
the case of ladders containing a single static hole. A simple
immediate application of this approach to classical ladders is
worked out.
Corresponding author: A. R. Pereira; 
e-mail: apereira@ufv.br; 
Tel.: +55-31-3899-2988, 
Fax: +55-31-3899-2483.

\newpage
\maketitle The interest in low dimensional quantum
antiferromagnets has been great ever since Haldane conjectured
\cite{Haldane83} that integer spin chains have a gap in their
excitation spectrum while half-integer spin chains do not. More
recently, spin ladders (two or more coupled spin chains) have also
attracted much interest, mainly when the effects upon doping are
considered. Indeed, when one manages to remove spins from the
system (leaving holes behind) the existence of superconductivity
is predicted \cite{Daggoto96} (and experimentally observed
\cite{Uehara96}). In this letter we would like to study the
presence of static holes in spin ladders with an arbitrary number
of legs by considering the nonlinear $\sigma$ model continum limit
of the model.

The continuum limit of pure antiferromagnetic Heisenberg spin
ladders with arbitrary number of legs has been derived by several
authors \cite{Ercolessi97,Sierra96,Senechal95,Ercolessi03}. Here
we apply the approach of Ref.\cite{Ercolessi97} for studying the
case of ladders containing static holes. We start with the
Hamiltonian for a pure ladder system with $n$ legs of length
$Na_{0}$ ($a_{0}$ is the lattice constant and $ N\gg n $) defined
as
\begin{eqnarray}
H=\sum_{a=1}^{n}\sum_{j=1}^{N}\left[J_{a}\vec{S}_{a}(j)\cdot
\vec{S}_{a}(j+1)+\tilde{J}_{a,a+1}\vec{S}_{a}(j)\cdot
\vec{S}_{a+1}(j)\right], \label{ladders}
\end{eqnarray}
where $\vec{S}_{a}(j)$ are the spin operators located in the $a$th
leg at the position $j=1,...,N$, while  $J_{a}>0$ and
$\tilde{J}_{a,a+1}>0$ are the antiferromagnetic exchange couplings
along the leg and rung respectively. The partition function of the
above Hamiltonian in the spin coherent state path-integral
representation is given by
\begin{eqnarray}
\textit{Z}(\beta)= \int [ \emph{D} \vec{\Omega}] \exp \left\{i
S\sum_{j,a}\omega [\vec{\Omega}_{a}(j,\tau)]-\int_{0}^{\beta} d
\tau H(\tau) \right\}, \label{partition}
\end{eqnarray}
where $\tau=it$ is the imaginary time variable, $\omega
[\vec{\Omega}_{a}(j,\tau)]$ is the Berry phase factor and
$H(\tau)$ is obtained by replacing the operator $\vec{S}_{a}(j)$
by the classical variable $S \vec{\Omega}_{a}(j,\tau)$ in the
Hamiltonian (\ref{ladders}). To get the continuum limit, it is
usual to assume that the dominant contribution to the path
integral comes from paths described by \cite{Haldane83,Haldane88}
\begin{eqnarray}
\vec{\Omega}_{a}(j,\tau)=(-1)^{a+j} \vec{\phi}(j,\tau)
\left(1-\frac{\mid \vec{l}_{a}(j,\tau) \mid^{2}}{S^{2}}
\right)^{1/2}+\frac{\vec{l}_{a}(j,\tau)}{S}. \label{Haldane}
\end{eqnarray}
The field $\vec{\phi}(j,\tau)$ is supposed to be slowly varying
and the fluctuation field $\vec{l}_{a}(j,\tau)$ is supposed to be
small ($\vec{l}_{a}(j,\tau)/S<<1$). The constraint
$\vec{\Omega}_{a}^{2}(j,\tau)=1$ implies
$\vec{\phi}^{2}(j,\tau)=1$ and $\vec{\phi}(j,\tau)\cdot
\vec{l}_{a}(j)=0$. Numerical works support the fact that the
staggered spin-spin correlation length is much greater than the
total width of the ladder \cite{Greven96,White94}. Then, assuming
that $\vec{\phi}(j,\tau)$ depends only on the site index $j$ along
the legs, Dell'Aringa \emph{et al.} \cite{Ercolessi97} mapped the
antiferromagnetic Heisenberg ladder system onto a (1+1) quantum
nonlinear $\sigma$ model
\begin{eqnarray}
Z_{\sigma} =\int  [ \emph{D}
\vec{\phi}]\exp(i\Gamma[\vec{\phi}])\exp \left\{-\frac{1}{2g}
\int_{0}^{\beta}d\tau\int dx \left[\frac{1}{v_{s}}(
\partial_{\tau} \vec{\phi})^{2}+
v_{s}(\partial_{x}\vec{\phi})^{2}\right]\right\}, \label{sigma1}
\end{eqnarray}
where $\Gamma[\vec{\phi}]=(\theta/4\pi) \int_{0}^{\beta}d\tau \int
dx \vec{\phi}\cdot (\partial_{\tau}\vec{\phi}\times
\partial_{x}\vec{\phi})$ (with $\theta=2 \pi
S$) for $n$ odd and $\Gamma[\vec{\phi}]=0$ for $n$ even,
reflecting the fact that, for half-spin systems, the excitation
spectrum has a gap (is gapless) when $n$ is even (odd). Besides,
the nonlinear $\sigma$ model parameters, the coupling constant $g$
and the spin wave velocity $v_{s}$, are defined by
\begin{eqnarray}
g^{-1}=S \left(\sum_{a,b,c}J_{a}L_{b,c}^{-1}\right)^{1/2},
\label{coupling1}
\end{eqnarray}
\begin{eqnarray}
v_{s}=S
\left(\frac{\sum_{a}J_{a}}{\sum_{b,c}L_{b,c}^{-1}}\right)^{1/2},
\label{velocity1}
\end{eqnarray}
where $L_{a,b}^{-1}$ is the inverse of the matrix
\begin{equation} \label{matexpr}
L_{a,b}=\left\{ \begin{array}{lll} 4J_{a}+\tilde{J}_{a,a+1}+ \tilde{J}_{a,a-1} &for& a=b\\
L_{a,b}=\tilde{J}_{a,a+1} & for & \mid a-b \mid= 1 \end{array} \right.
\end{equation}
with $\tilde{J}_{a,a+1}\equiv \tilde{J}_{a-1,a}$  and $\tilde{J}_{1,0}=\tilde{J}_{n,n+1}=0$.

Now we consider the system in the presence of static holes (spins
removed from the ladder). In two spatial dimensions, one of the
simplest way of studying this problem in the continuum limit is
through a non-simply connected manifold
\cite{Pereira+05,Paula+04,Pereira06}. In this case a disk is
removed from the magnetic plane, leaving a hole behind, and this
hole is interpreted as a nonmagnetic impurity (or a spin vacancy)
since there is no magnetic degrees of freedom insight it. This
approach has good qualitative and quantitative agreement with
numerical calculations
\cite{Pereira+05,Paula+04,Pereira06,Paula05}. However, in the case
of ladders as described by the nonlinear $\sigma$ model given by
Eq. (\ref{sigma1}), we cannot simply remove a disc from the space
because the problem becomes essentially one-dimensional. There is
no possibility of removing a part of the space without breaking
the ``effective" lattice. Then, within the above approach a
``hole" must affect the exchange interactions $J_{a}$ and
$\tilde{J}_{a,a\pm 1}$. As a consequence,  the matrix $L_{a,b}$
becomes $j$-dependent. In such a way that the lattice is not
broken for a single defect. It means that the parameters $g$ and
$v_{s}$ are now functions of the position along the ladder.
Depending on the number of legs, there are more than one position
to put a single vacancy which yields different results. Some
examples are shown in  Fig.(1).

With the above considerations in mind, our approach for the
nonlinear $\sigma$ model describing spin ladders with a static
hole centered at $j=x_{0}$ (along the legs), $a=k$ (along the
rungs) are summarized as follows
\begin{eqnarray}
Z_{\sigma,hole} =\int  [ \emph{D}
\vec{\phi}]\exp(i\Gamma[\vec{\phi}])\exp \left(-\int_{0}^{\beta}d\tau\int dx
{\cal L}_{\sigma,hole}\right), \label{sigmahole}
\end{eqnarray}
with
\begin{eqnarray}
{\cal L}_{\sigma,hole} =& \int dx \frac{1}{2g_{k}(x-x_{0})}
\left[\frac{1}{v_{s,k}(x-x_{0})}(
\partial_{\tau} \vec{\phi} )^{2}+
v_{s,k}(x-x_{0})(\partial_{x}\vec{\phi})^{2}\right].
\label{Hamiltoniansigma}
\end{eqnarray}
In order to explicitly define the new parameters space dependents
$g_{k}(x-x_{0})$ and $v_{s,k}(x-x_{0})$, we first rewrite the
$n\times n$ matrix $L_{a,b}$ as follows
\begin{eqnarray}\label{Matrix1}
L_{a,b}=\left(
  \begin{array}{ccccccccccc}
     L_{1,1} & L_{1,2} & 0 & 0 & 0 &... & 0 & 0 & 0 & ...& 0  \\
     L_{2,1} & L_{2,2} & L_{2,3} & 0 &0& ...& 0 & 0 & 0 &...& 0 \\
     0 & L_{3,2} & L_{3,3} & L_{3,4} & 0&...& 0 & 0 & 0 &...& 0 \\
     \vdots & \vdots & \vdots & \vdots & \vdots & \ddots & \vdots
     & \vdots & \vdots & \ddots & \vdots \\
     0 & 0 & 0 & 0 & 0& ...& L_{k-2,k-1} & 0 & 0 &...& 0\\
      0 & 0 & 0 & 0 & 0& ...& L_{k-1,k-1} & L_{k-1,k} & 0 &...& 0 \\
      0 & 0 & 0 & 0 & 0& ...& L_{k,k-1} & L_{k,k} & L_{k,k+1} &...&0 \\
      0 & 0 & 0 & 0 & 0& ...& 0 & L_{k+1,k} & L_{k+1,k+1} &...&0 \\
      0 & 0 & 0 & 0 & 0& ...& 0 & 0 & L_{k+2,k+1} &...&0 \\
      \vdots & \vdots & \vdots & \vdots & \vdots & \ddots & \vdots
     & \vdots & \vdots & \ddots & \vdots \\
     0 & 0 & 0 & 0 & 0 & ...& 0 & 0 & 0 &...&L_{n,n}\\
  \end{array}
\right).
\end{eqnarray} \\
If we place the vacancy  at $(j,a)=(x_{0},k)$, the above matrix
will be the same for all $j \neq x_0, x_0-1$ and equal to
(\ref{matexpr}). For $j=x_0-1$ the matrix will be again of rank
$n$ with a slightly different coefficient $L_{k,k}$. For $j=x_0$
the matrix has zeroes along the $k-th$ row and the $k-th$ column.
Thus we define a new matrix $K_{a,b}$ of order $(n-1) \times
(n-1)$, which has almost the same elements of the above matrix and
without line $k$ and column $k$. Explicitly:
\begin{eqnarray}\label{Matrix2}
K_{a,b}=\left(
  \begin{array}{cccccccccc}
     L_{1,1} & L_{1,2} & 0 & 0 & 0 &... & 0 &  0 & ...& 0  \\
     L_{2,1} & L_{2,2} & L_{2,3} & 0 &0& ...& 0 &  0 &...& 0 \\
     0 & L_{3,2} & L_{3,3} & L_{3,4} & 0&...& 0 &  0 &...& 0 \\
     \vdots & \vdots & \vdots & \vdots & \vdots & \ddots & \vdots
     & \vdots & \ddots & \vdots \\
     0 & 0 & 0 & 0 & 0& ...& L_{k-2,k-1} &  0 &...& 0\\
      0 & 0 & 0 & 0 & 0& ...& K_{k-1,k-1} & 0 &...& 0 \\
      0 & 0 & 0 & 0 & 0& ...& 0 & K_{k+1,k+1} &...&0 \\
      0 & 0 & 0 & 0 & 0& ...& 0 & L_{k+2,k+1} &...&0 \\
      \vdots & \vdots & \vdots & \vdots & \vdots & \ddots & \vdots
      & \vdots & \ddots & \vdots \\
     0 & 0 & 0 & 0 & 0 & ...& 0  & 0 &...&L_{n,n}\\
  \end{array}
\right),
\end{eqnarray} \\
where $K_{k-1,k-1}=L_{k-1,k-1}-\tilde{J}_{k-1,k}$ and
$K_{k+1,k+1}= L_{k+1,k+1}-\tilde{J}_{k+1,k}$. This formula holds
only in the region of the spin vacancy ($\mid x-x_{0}\mid \lesssim
a_{0}$). Therefore we may assume that the coefficients $g$ and $v$
are given by:

\begin{equation}\label{coupling2}
g_{k}^{-1}(x-x_{0})= \left\{
\begin{array}{lc}
g^{-1}  & {\rm \emph{for}} \; \mid x-x_{0}\mid \gtrsim a_{0} , \\
S \left(\sum_{a\neq k,b,c}J_{a}K_{b,c}^{-1}\right)^{1/2} & {\rm
\emph{for}} \; \mid x-x_{0} \mid \lesssim a_{0},
\end{array}
\right.
\end{equation}

\begin{equation}\label{velocity2}
v_{s,k}(x-x_{0})= \left\{
\begin{array}{lc}
v_{s}  & {\rm \emph{for}} \; \mid x-x_{0}\mid \gtrsim a_{0} , \\
S \left(\frac{\sum_{a\neq
k}J_{a}}{\sum_{b,c}K_{b,c}^{-1}}\right)^{1/2} & {\rm \emph{for}}
\; \mid x-x_{0} \mid \lesssim a_{0}.
\end{array}
\right.
\end{equation}

The change in the parameters $g_{k}(x-x_{0})$ and
$v_{k,s}(x-x_{0})$ in the zone of influence of the vacancy is
associated with the discontinuous  change in the number of legs in
this region (see Fig.(1)). Of course, the field $\vec{\phi}$ must
be continuous across the pure and impure regions. Note that, in
principle, only the coupling constant and spin wave velocity are
(locally) affected by the removed spin. As the Berry phases do not
depend on these values, they are not very sensitive to the
presence of the defect (see Eq. (\ref{sigmahole})). It means that
the ground state may not be very affected by the presence of the
impurity. These results are in agreement with recent numerical
calculations\cite{Anfuso06}, which give evidences that the ground
state configuration of the entire ladder system is not changed
significantly by the impurity except for the local extraction of
the missing bonds. Indeed, the energy cost to remove a spin from a
two-leg ladder with spin-1/2 and $J=\tilde{J}$ is $E\approx
1.215J$ \cite{Anfuso06}, which is almost completely accounted for
by the missing energy bonds along the legs ($0.350J$) and across
the rungs ($0.455J$). Then, the method developed here is a good
approximation and can be generalized for ladders containing a low
concentration of impurities. Below we give a simple application of
this approach. Our example is done for classical spin systems
because the calculations are almost direct in this case. Besides,
there are also manganese halide compounds \cite{Jongh74} that are
quasi-one-dimensional and two-dimensional antiferromagnets.
Furthermore, these $Mn(II)$ compounds have spin $5/2$ so they are
also nearly classical and therefore, potential systems to test our
results.

At zero temperature, the Hamiltonian of spin ladders possesses, in
the classical limit ($S \rightarrow \infty$), a minimum given by
the antiferromagnetic vacuum solution $\vec{\phi}_{0}(x)=
\phi_{z}\hat{z}= \hat{z}$, where $\hat{z}$ is an unit vector in
the vertical direction. This solution breaks the $O(3)$ invariance
of the model down to the subgroup $O(2)$ of rotations around the
z-axis. Consequently there should appear two Goldstone modes,
which are nothing but spin waves, associated with $\phi_{x}$ and
$\phi_{y}$. Then a natural first step concerning the impurity
systems is to study the interactions between spin waves and holes.
In a quantum spin system, the corresponding problem would be the
interactions between triplons \cite{Notbohm06} (which are a
triplet of well defined spin-1 magnons) and holes. However, it
will be considered in a future work.

Expanding $\vec{\phi}(x,t)$ around the vacuum solution $
\vec{\phi}(x,t)=\vec{\phi}_{0}(x)+\vec{\eta}(x,t)$ and minimizing
Hamiltonian (\ref{Hamiltoniansigma}), one obtains, in the
linearized approximation, the following scattering equation
$\partial_{x}^{2}\vec{\eta}(x,t)-[1/v_{s,k}^{2}(x-x_{0})]\partial_{t}^{2}\vec{\eta}(x,t)=U_{k}(x)\vec{\eta}(x,t)$,
where the scattering potential is $U_{k}(x)=-\{\partial_{x}\ln
[v_{s,k}(x-x_{0})/g_{k}(x-x_{0})]\}\partial_{x}$. However, the
function $v_{s,k}(x-x_{0})/g_{k}(x-x_{0})$ is constant practically
through all space (see Eqs.(\ref{coupling2})) while
(\ref{velocity2})  varies only when entering  the impurity
regions. Therefore, $U_{k}(x)$ is zero in almost all space and we
have magnon solutions for the field equation in the three regions:
($x<x_{0}-a_{0}$), ($x_{0}-a_{0} <x<x_{0}+a_{0}$) and
($x>x_{0}+a_{0}$). The form of $U_{k}(x)$ is not explicitly known
but its effects can be envisaged using the following simple
analysis: if a magnon (for instance, coming from the left) hits
the zone of influence of the potential $U_{k}(x)$ (or the zone of
the impurity) at $x_{0}-a_{0}$, then its velocity will be changed
from $v_{s}$ to $v_{k,s}=(\sum_{a\neq
k}J_{a}/\sum_{b,c}K_{b,c}^{-1})^{1/2}$ and after leaving behind
this region at $x_{0}+a_{0}$, it will be changed again to $v_{s}$.
Consequently, supposing a plane wave coming from $-\infty$, and
assuming that the lowest order effect of $U_{k}(x)$ is to cause
elastic scattering centers for magnons, the solution at $+\infty$
can be approximated by $\vec{\eta}_{0}\exp[i(qx-\omega_{q}
t+\delta_{n,k}(q)/2)]$ with frequency $\omega_{q}=qv_{s}$, which
of course, is precisely the dispersion relation for magnons in the
absence of holes. The function $\delta_{n,k}(q)$ may be regarded
as a phase-shift (for magnons in a ladder with $n$ legs) which
depends on the particular position (leg $k$) of the hole in the
spin ladder for a determined wave number $q$. Indeed, after
passing the defect, the wave is shifted from the original one due
to the different velocity $v_{k,s}$ acquired in the hole region.
This phase-shift can be easily estimated considering the
difference of paths $v_{s}\Delta t$ and $v_{k,s}\Delta t$, where
the interval $\Delta t=2a_{0}/v_{k,s}$ is the time necessary for
the wave to leave behind the region of the hole ($x_{0}-a_{0} <x
<x_{0}+a_{0}$). In the lowest order, it is given by
\begin{eqnarray}
\delta_{n,k}(q)=-4qa_{0} \left(\frac{v_{s}}{v_{k,s}}-1
\right)=-4qa_{0} \left[\left(\frac{\sum_{a}J_{a}}{\sum_{a\neq
k}J_{a}}\right)^{1/2}\left(\frac{\sum_{b,c}K_{b,c}^{-1}}{\sum_{b,c}L_{b,c}^{-1}}\right)^{1/2}-1
\right].
\end{eqnarray}

For the case of a spin ladder with two legs and $N\rightarrow
\infty$, the phase-shift does not depend on the position of the
vacancy, which can be put at the left ($k=1$) or right ($k=2$)
leg. For $J_{1}={J_{2}}$ and $\tilde{J}_{a,b}=\tilde{J}$, it is
given by
\begin{eqnarray}
\delta_{2,1}(q)=\delta_{2,2}(q)= -4qa_{0}
[(1+\tilde{J}/2J)^{1/2}-1] .
\end{eqnarray}
Spin ladders with three or more legs have two different
situations. For example, for the case with three legs, if the
vacancy is placed at the first or last leg, we have:
\begin{eqnarray}
\delta_{3,1}(q)=\delta_{3,3}(q)= -4qa_{0}
\left\{\left[\frac{(1+3R/4)}{(1+R/2)(1+R/12)}
\right]^{1/2}-1 \right\},
\end{eqnarray}
where $R=\tilde{J}/J$. By the other hand, if the hole is placed at
the central leg, the phase shift is
\begin{eqnarray}
\delta_{3,2}(q)= -4qa_{0}
\left\{\left[\frac{(1+3R/4)}{(1+R/4)(1+R/12)}
\right]^{1/2}-1 \right\}.
\end{eqnarray}

Finally we discuss the magnon density of states in the impurity
spin ladders. To do this, we consider a large system of size
$Na_{0}$ and impose periodic boundary conditions on the continuum
(magnons) states $\vec{\eta}_{q}(x)$. This periodicity, together
with $\vec{\eta}_{q}(x)_{x\rightarrow \pm \infty}\approx
\vec{\eta}_{0}\exp[i(qx\pm\delta_{n,k}(q)/2)]$, gives the
following condition for the allowed wave vectors:
$Na_{0}q_{m}+\delta_{n,k}(q_{m})=2\pi m$ ($m=0,\pm 1, \pm 2,...$).
Clearly, the magnon density of states is changed by the presence
of a hole as follows
\begin{eqnarray}
\varrho_{n,k}(q)=\frac{d
m}{dq}=\frac{Na_{0}}{2\pi}+\frac{1}{2\pi}\frac{d
\delta_{n,k}(q)}{dq},
\end{eqnarray}
and so, the change in the density of states $\Delta
\varrho_{n,k}(q)=\varrho_{n,k}(q)-Na_{0}/2\pi$ is given by
\begin{eqnarray}
\Delta \varrho_{n,k}(q)=\frac{1}{\pi}\frac{d \delta_{n,k}(q)}{dq}.
\end{eqnarray}
Above we have multiplied by a factor of 2 to take into account the
two Goldstone modes. For a classical two-leg spin ladder one has
\begin{eqnarray}
\Delta \varrho_{2,1}(q)=\Delta
\varrho_{2,2}(q)=\frac{-4a_{0}}{\pi} [(1+R/2)^{1/2}-1].
\end{eqnarray}
For three-leg ladder
\begin{eqnarray}
\Delta \varrho_{3,1}(q)=\Delta \varrho_{3,3}(q)=
\frac{-4a_{0}}{\pi}
\left\{\left[\frac{(1+3R/4)}{(1+R/2)(1+R/12)}
\right]^{1/2}-1 \right\}
\end{eqnarray}
and
\begin{eqnarray}
\Delta \varrho_{3,2}(q)= \frac{-4a_{0}}{\pi}
\left\{\left[\frac{(1+3R/4)}{(1+R/4)(1+R/12)}\right]^{1/2}-1
\right\}.
\end{eqnarray}

In Fig.(2) we plot $D=\Delta \varrho_{n,k}/a_{0}$ as a function of
$R=\tilde{J}/J$ for $n=2$ and $n=3$. In general, the density of
states does not depend on the wave-vector $q$ (since the
phase-shift has a linear dependence on $q$). For all legs, the
limit $R\rightarrow 0$ implies $D\rightarrow 0$. As expected, it
means that the presence of $\tilde{J}$ is very important for the
phase-shifts as well as for the change in the density of states.
In practice, it avoids a broken lattice and leads to $v_{k,s}\neq
0$. For a three-leg ladder with the impurity placed at leg 1 or 3,
$\Delta \varrho_{3,1}$ is very small for an appreciable range of
$R$ and becomes positive for $R>4$. It increases considerably for
large values of $R$ and in the limit $R\rightarrow \infty$,
$\Delta \varrho_{3,1}\rightarrow 1.27$. Such a general behavior is
also expected for ladders with more than three legs with
impurities placed at the external legs (of course, it may have
important qualitative changes as, for example, the function is
positive for $R<R_{c}$ and then becomes positive for $R>R_{c}$).
If the vacancy is located at the central leg, the behavior of the
three-leg ladder is similar to the previous case but the change in
the density of states is much larger. In this case, $\Delta
\varrho_{3,2}(q)/a_{0}$ becomes positive only for $R>20$ (not
shown in Fig.2). In addition , like $\Delta
\varrho_{3,1}(q)/a_{0}$, $\Delta \varrho_{3,2}(q)/a_{0}\rightarrow
1.27$ as $R\rightarrow\infty$. On the other hand, for a two-leg
ladder, $\Delta \varrho_{2,1}(q)/a_{0}$ is always negative and its
modulus increases monotonically as $R$ increases. The magnon
density of states suffers a very expressive change for large
values of $\tilde{J}$. Therefore, in this circumstance, there is a
clear distinction in the magnon spectrum when a hole in the spin
ladder is present or absent.

Ladders with a higher number of legs might be approached as well,
using for example the Mathematica program. In particular it is
interesting to study how the ratio $w= v_s/v_{k,s}$ changes as we
increase the number $n$ of legs and approach a two-dimensional
lattice. We have checked this for the isotropic case $J =
\tilde{J}$ by putting the hole in the center leg of the ladder,
getting the following results for $n=11, 31, 51,71,91$
respectively: $w = 1.03358, 1.01029, 1.00605, 1.00428, 1.00331$.
One can immediately see that $w$ approaches zero as $n$ goes to
infinity. However, the method used here for the phase-shifts is
not convenient for two-dimensional (2d) systems because one has to
deal with cylindrical waves, which contains an infinite number of
angular momentum channels. A more adequate method for the 2d case
in the continuum approximation is given in Ref.\cite{Pereira06}.
Thus, we cannot conclude that the influence of a single static
hole on the magnon phase shift and density of states becomes
negligible in the two-dimensional system.

We also expect that a similar behavior may happen to triplons in
spin-1/2 two-leg ladders. Basic differences must appear due to the
existence of a gap for these last excitations. Of course, the
change in the spin wave density of states has a deep influence on
the static and dynamical properties of ladders containing a low
concentration of holes as it causes a change in the magnon free
energy \cite{Currie80}.

In summary, we have proposed an approach based on the nonlinear
$\sigma$ model to study static holes in spin ladder systems with
an arbitrary number of legs. As an immediate example of
application of this method, we have studied the magnon-hole
interactions in classical spin ladders and calculated the magnon
phase-shifts and the consequent change in the magnon density of
states. Such results may also give some insights about the
behavior of linearized oscillatory excitations around holes in
low-dimensional quantum spin materials. However, the calculations
were performed using a long wavelength theory and we have to
remark that in experimental situations involving small scales (of
order of the lattice size) the application of the nonlinear
$\sigma$ model should be viewed with caution. Future study
considering systems containing multiple impurities (which, for
adjacent holes, have larger sizes) and applications to quantum
ladders are in progress. As in the two-dimensional case
\cite{Paula+04,Pereira06}, our results may also be useful to study
possible topological excitations interacting with holes in spin
ladders.

\newpage

\centerline{\large\bf Acknowledgements} \vskip .3cm The authors
thank CAPES (Brazilian agency) for financial support. This work
was partially supported
by CNPq (Brazil), the TMR network EUCLID and the Italian MIUR through COFIN projects.\\

\newpage
Figure Captions

Figure 1. Configurations of spin ladders along the z-direction
with two and three legs containing a hole. For the case of
three-leg ladders, the second and third configurations are
equivalents, but the fourth leads to different results.

Figure 2. Change in the density of states $D=\Delta
\varrho_{n,k}/a_{0}$ as a function of the rung coupling
$R=\tilde{J}/J$ for spin ladders with two-leg (solid line) and for
the two possibilities for three-leg: vacancy placed at the central
leg (dashed line) and at leg 1 or 3 (dotted line).

\clearpage \hspace{40cm}
\begin{figure}
\begin{center}\resizebox{12cm}{!}{\includegraphics{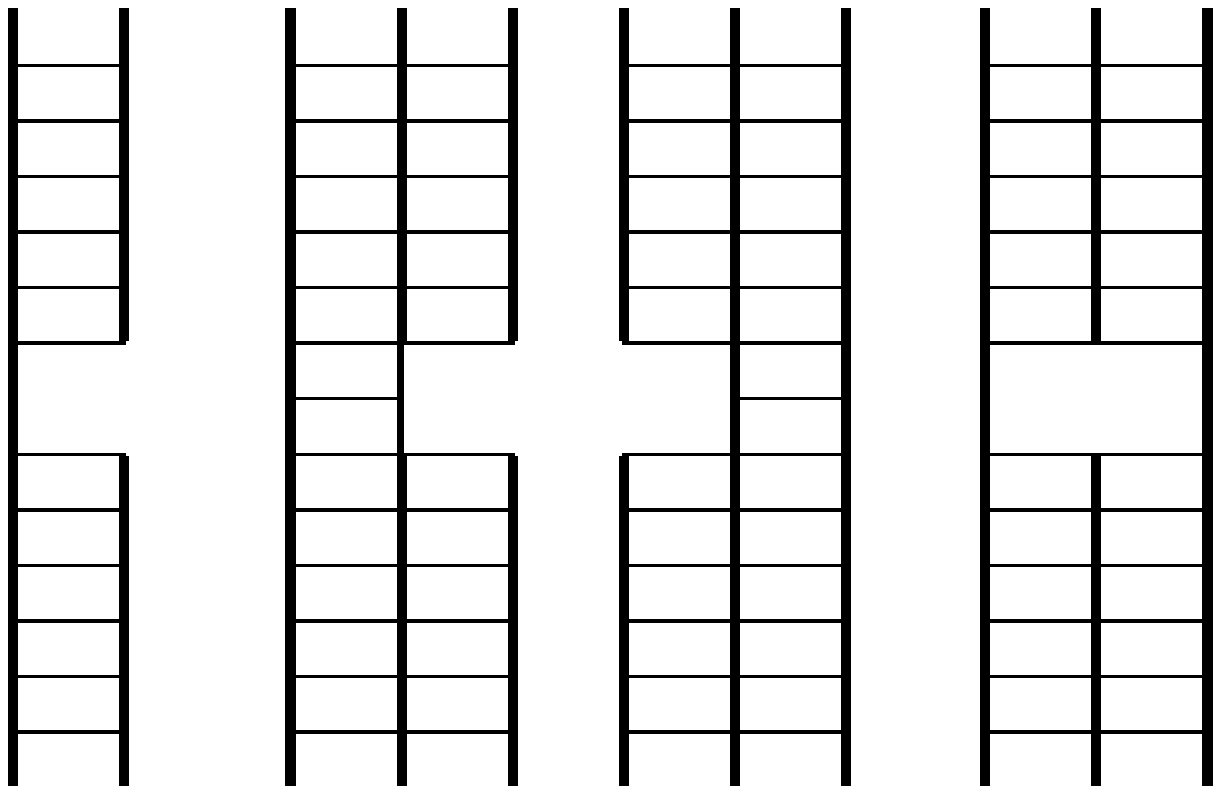}}\end{center}
\end{figure}
\clearpage

\clearpage \hspace{40cm}
\begin{figure}
\begin{center}\resizebox{12cm}{!}{\includegraphics{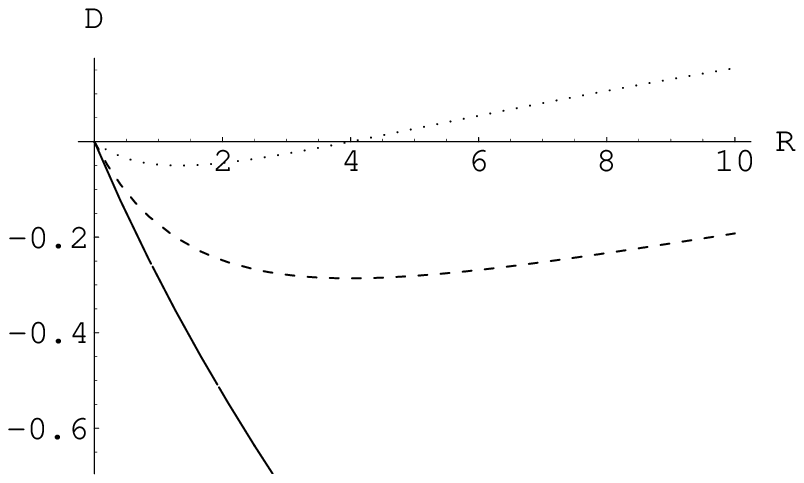}}\end{center}
\end{figure}
\clearpage

\end{document}